\documentstyle[psfig,aps,twocolumn]{revtex} 

\begin{document}
\draft 
\title{Band Population Measurements\\in a\\Purely Optical Dark Lattice}
\author{Frank  Sander, Tilman Esslinger, and Theodor~W.~H{\"a}nsch}
\address{Sektion Physik, Ludwig-Maximilians Universit{\"a}t M{\"u}nchen, 
D-80799 Munich, Germany and\\
Max-Planck-Institut f{\"u}r Quantenoptik, D-85748 Garching, Germany}
\author{Herwig Stecher and Helmut Ritsch}
\address{Institute for Theoretical Physics, University of Innsbruck, 
A-6020 Innsbruck, Austria} 

\date{\today} 
\maketitle

\begin{abstract}\vspace*{-1.7\baselineskip}

\noindent We create a dark optical lattice structure using a blue
detuned laser field coupling an atomic ground state of total angular
momentum $F$ simultaneously to two excited states with angular momenta
$F$ and $F-1$, or $F$ and $F+1$.  The atoms are trapped at locations
of purely circular polarization.  The cooling process efficiently
accumulates almost half of the atomic population in the lowest energy
band which is only weakly coupled to the light field.  The populations
of the two lowest energy bands reaches 70\%. Kinetic energies on the
order of the recoil energy are obtained by adiabatically reducing the
optical potential.  The band populations are directly mapped on free
particle momentum intervals by this adiabatic release. In an
experiment with subrecoil momentum resolution we measure the band
populations and find good absolute agreement with the theoretically
calculated steady state band populations.
\end{abstract}

\pacs{laser cooling and trapping, atomic physics in general}

\section{Introduction}

\noindent
Optical lattices present a unique quantum-mechanical system for the
study of matter waves interacting with a perfectly periodic
potential. The periodic potential is formed by the light shift that an
atom experiences in a near-resonant standing wave. The atoms can be
cooled to the minima of the optical potential forming a 
periodic array of microscopic traps\cite{Prentiss}.  It is a
fascinating question whether atom-atom-interactions
\cite{AtomAtomInteractions} and quantum statistical effects, similar to
those observed in magnetic traps\cite{BEC}, can be studied in optical
lattices when two or more atoms occupy a single lattice site. This
requires a technique for achieving much higher atomic densities than
possible in optical lattices at present. To overcome the density limit
that is due to light induced interactions between atoms optical
lattices have been proposed \cite{Courtois,DarkLatticeProposal}, in which the
fluorescence rate for the trapped atoms is strongly reduced. They are
referred to as dark or gray optical lattices and have been
demonstrated \cite{Hemmerich} using a scheme that requires
a static magnetic field.

In this paper we describe and experimentally investigate a new type of
dark (or gray) optical lattice that utilizes merely an optical field.
The lattice combines efficient accumulation of the atoms in the lowest
energy level of the optical potential with a reduced fluorescence rate
for the localized atoms. We have studied \cite{OL,PRA} the adiabatic release of
atoms from the dark optical lattice both theoretically and
experimentally with the result that the band populations are indeed
mapped on the corresponding momentum intervals of the free atom, as it
was suggested by Kastberg {\sl et al.}  \cite{Philipps95} for
adiabatic cooling in a bright optical lattice. This population mapping
is accurate, if optical pumping between different bands is negligible
during the adiabatic release. In an experiment with a slow beam of Rb
atoms we utilize this population mapping to measure the band populations
of the purely optical dark lattice.  For a wide range of parameters we
find excellent agreement between the measured band populations and
theoretically calculated values. This establishes population mapping
by adiabatic release as an experimental tool to investigate the
interaction of atoms with a spatially periodic potential.

\section{Purely Optical Dark Lattice}

\begin{figure}[h]
\centerline{\psfig{file=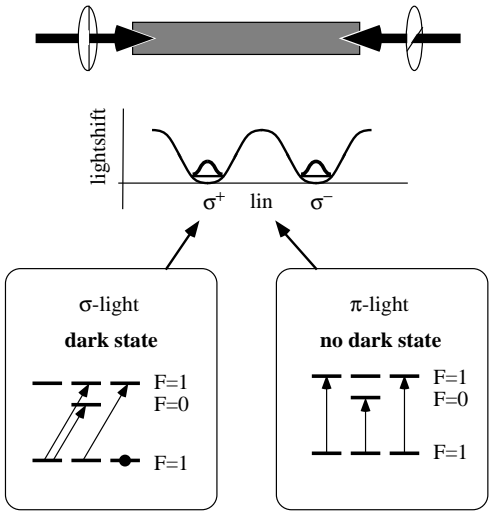,angle=0,width=0.36\textwidth}}
\caption{The purely optical dark lattice on a $F=1\to 1,0$
transition pair. The atoms are accumulated at locations of pure
$\sigma$ polarization in the standing wave, where they largely
decouple from the light field\protect\cite{QuantizationAxis}.}
\label{Lattice}
\end{figure}

\noindent
To create the optical lattice we use a standing wave consisting of two
counter propagating waves with mutually orthogonal linear polarizations
(lin$\perp$lin).  In this configuration the polarization changes
periodically from $\sigma$ to $\pi$ polarization along the standing
wave axis. Detuning and intensity are chosen such, that the laser
field couples a ground state manifold with integer total angular
momentum $F$ to two excited state manifolds with angular momenta $F$
and $F-1$.  Fig.\ \ref{Lattice} illustrates the situation for the
simplest example of a $F=1\to F=1,0$ transition pair.  The interaction
of the oscillating atomic dipole with the standing wave causes
spatially periodic light shifts in the atomic ground state
manifold. In regions of purely $\sigma^+$ ($\sigma^-$) polarized light
the atoms are optically pumped into the $m_F$=$1$ ($m_F=-1$) ground
state, which is decoupled from the light field and experiences no
light shift or optical excitation. At locations of linearly polarized
light all ground state sublevels are coupled to excited
states\cite{QuantizationAxis}. In this case all ground state sublevels
are light shifted towards higher energies, if the laser field is tuned
to a frequency higher than both transition frequencies (blue
detuning). In this semi classical picture of point like atoms one
expects that the atoms are cooled by a Sisyphus
mechanism\cite{Sisyphus} and accumulated in dark states at locations
of pure $\sigma$-polarization. In a picture which treats the atomic
motion quantum mechanically the atomic wave function always has a
finite spatial extent and can not be completely decoupled from the
light field. Therefore a dark (gray) optical lattice with a low
fluorescence rate and long coherence times is formed \cite{PRA}.  Low
fluorescence rates for the localized atoms are achieved, even if the
lattice is operated at small detunings. The situation is qualitatively
similar to that of magnetic field induced dark optical lattices, where
localized dark states are created by combining a standing wave with a
magnetic field \cite{Courtois,Hemmerich}.

We have also investigated a second configuration\cite{OL}, in which a
purely optical dark lattice is formed in a similar way on a $F\to
F,F+1$ transition pair with integer angular momentum $F$. The laser
field forms a lin$\perp$lin standing wave and is tuned between the two
transitions.  To understand the underlying mechanism consider first
the coupling of the laser field only to the $F\to F$ transition. For
any polarization and for any position in space one substate
$\psi_{nc}$ of the ground state manifold is not coupled to the light
field.  All other substates of the ground state manifold couple to the
light and are light shifted towards higher energies, because the laser
field is tuned to a frequency higher than the transition
frequency. Now complete the picture and consider additionally the
coupling of the laser field to the $F+1$ excited state manifold. Here
the laser field is detuned towards lower frequency and introduces
additional spatially varying light shifts towards lower energies. This
affects all substates of the ground state manifold. The state
$\psi_{nc}$, which so far was unshifted, undergoes the largest light
shift towards lower energy. As a result a dark optical lattice is
formed, since optical pumping on the (far detuned) $F\to F+1$
transition is negligible.  In spite of low optical pumping dynamics on
the $F\to F+1$ transition the coupling to the $F\to F$ transition
allows for an efficient Sisyphus cooling\cite{Sisyphus} of the atoms
to the minima of the lowest optical potential.

\section{Adiabatic Mapping}

\begin{figure}[h]
\centerline{\psfig{file=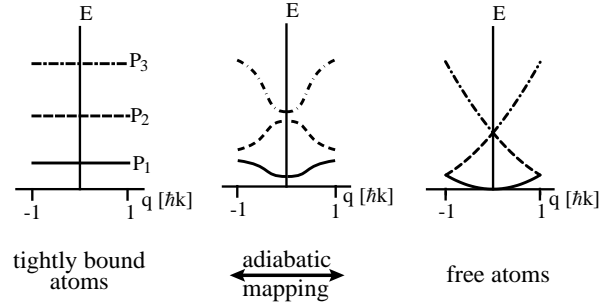,angle=0,width=0.45\textwidth}}
\caption{Band structure during the adiabatic
release. Vertical axis: energy of the Bloch states. Horizontal axis:
Bloch index. The three lines correspond to the three lowest energy bands.}
\label{LatticeRelease}
\end{figure}

\noindent
It has been suggested in recent experimental work\cite{Philipps95,OL}
that the populations of an optical lattice can be directly
experimentally measured by an adiabatic release of the atoms from the
lattice and a subsequent measurement of the resulting atomic momentum
distribution. Fig.~\ref{LatticeRelease} shows schematically the band
structure of the lattice during the adiabatic release. The plot on the
left side corresponds to a situation of high laser field intensity,
{\sl i.~e.~}of deep optical potentials. The atoms are tightly bound to
the lattice sites and tunneling between neighboring potential wells
is negligible. Therefore the lowest energy bands are
flat\cite{AshcroftMurmin}. For lower field intensity (central plot)
the wave functions of neighboring wells can overlap and the energy
bands have a corresponding curvature. If the field is completely
switched off and no lattice interaction is present, the band structure
is that of a free particle. The segments of the
energy-momentum-parabola $E=P^2/(2M)$ are plotted on the right side in the
reduced zone scheme. If the release is fully adiabatic, the population
from the lowest (first) band is mapped exactly to the momentum
interval between $-\hbar k$ and $+\hbar k$ and the second band will be
mapped to the intervals $-2\hbar k$ to $-\hbar k$ and $\hbar k$ to
$2\hbar k$. The $n$-th band will be mapped on the $-n\hbar k$ to
$-(n-1)\hbar k$ and the $(n-1)\hbar k$ to $n\hbar k$
momentum intervals. Nonadiabaticity and incoherent redistribution of the atoms
during the release will modify this mapping. The assignment between the
band populations of the lattice and the momentum intervals of the free
atoms will be less accurate.  We have performed a full quantum Monte
Carlo simulation with a time dependent laser intensity to
quantitatively verify the mapping between the stationary population
distribution of the lattice and the momentum distribution of the free
atom. For the three lowest energy bands the deviations between the
band populations and the calculated fraction of atoms to be found in
the corresponding momentum interval are below 1\% \cite{PRA}.

\section{Experiment}

\noindent
For an experimental demonstration of the purely optical dark lattice
and the band population measurement, we have performed an experiment
\cite{PRA} with the apparatus described in
Refs.\cite{Esslinger,OL}. The purely optical dark lattice is realized
on the $F=3 \to F=3,2$ transitions of the $^{85}$Rb $D_1$-line.

\begin{figure}[h]
\centerline{\psfig{file=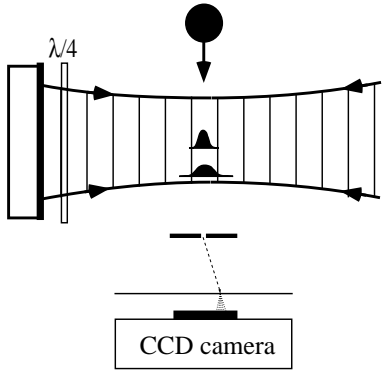,angle=0,width=0.28\textwidth}}
\caption{Experimental setup. A cold beam of Rb atoms crosses the
standing wave. Due to the Gaussian beam profile the atoms are
adiabatically released from the optical lattice. The
momentum distribution is detected below the standing wave. See text.}
\label{ExperimentSetup}
\end{figure}

A pulsed beam of cold Rubidium atoms is directed vertically downwards
and crosses a standing wave field (lattice field), which induces the
optical potentials of the dark lattice (Fig.~\ref{ExperimentSetup}).
The atoms are cooled into the lattice sites and are then gradually
released from the optical potential due to the Gaussian shape of the
lattice field.  Below the lattice field the momentum distribution of
the atoms is measured with a resolution of one third of the photon
recoil.

The lattice field is induced by a standing wave oriented along the
$x$-axis and has a frequency tuned
$\Delta_3=26\,\Gamma=2\pi\cdot145\,$MHz to the blue of the $F=3 \to
F=3$ transition. The hyperfine-splitting between the two excited
states of the $D_1$-line is $65\,\Gamma$. Correspondingly the detuning
with reference to the $F=3\to F=2$ transition is $\Delta_{2} =
91\,\Gamma$.  The incoming beam of the standing wave is linearly
polarized along the $z$-axis and the reflected beam is polarized along
the y-axis.  The Gaussian waists of the beams are $w_z=1.35$\,mm in
$z$-direction and $w_y=0.48$\,mm in $y$-direction.  This corresponds
to a 0.4\,ms time of flight of the atoms (3.2\,m/s) through the waist
$w_z$.  The region of the lattice field is shielded against stray
magnetic fields to well below 0.5\,mG.  A second standing wave
overlaps the lattice field.  It is tuned to the $F=2 \to F=3$
transition of the $D_2$-line and optically pumps the atoms into the
$F=3$ ground state.

To determine the momentum distribution we place a pinhole of
$75\,\mu$m diameter 5\,mm below the standing wave axis. The atoms
that pass through the pinhole expand
horizontally in two dimensions according to their transversal
momentum. A transversal momentum of $1\,\hbar k$ is translated into
$170\,\mu$m transversal displacement in a plane 9.6\,cm below the
pinhole.  The spatial distribution of the atoms in this plane is
imaged by recording the fluorescence in a sheet of light with a CCD
camera.  The sheet of light is formed by a standing wave which is
resonant with the $F=3 \to F=4$ closed cycle transition of the
$D_2$-line.  For each set of parameters we accumulate 200 single shot
images and subtract the separately measured background.  To obtain a
one dimensional momentum distribution in $x$-direction we integrate
the two dimensional distributions along the $y$-axis.

\begin{figure}[h]
\centerline {\psfig{file=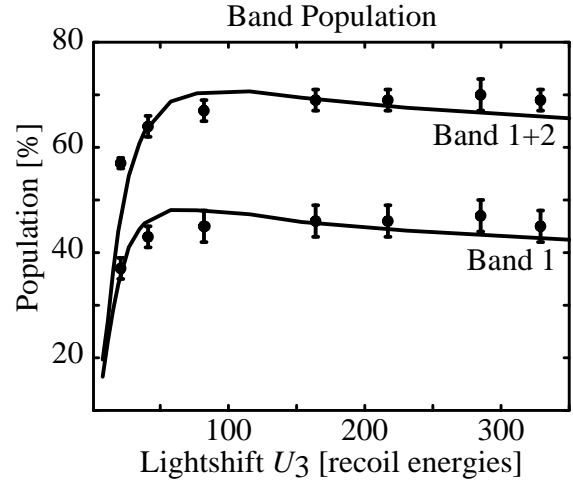,width=0.43\textwidth}}
\caption {Measured populations (data points) of the lowest and of the
two lowest bands of the optical lattice as a function of the
light shift $U_3$ induced by the $F=3 \to F=3$ transition. The solid
lines are the steady state result of a rate equation calculation for
the center of beam parameters.}
\label{popula}
\end{figure}

We count the number of atoms detected in the momentum intervals
$-\hbar k$ to $+\hbar k$ and $-2\hbar k$ to $+2\hbar k$ corresponding
to the populations in the lowest energy band and the two lowest energy
bands, respectively.  These experimentally obtained band populations
are plotted as data points in Fig.~\ref{popula}. The light shift is
given by $U_3=s_3\Delta_3((2\Delta_3/\Gamma)^2+1)^{-1}/2$, where $s_3$
is the resonant saturation parameter on the $F=3 \to F=3$ transition
in the center of the Gaussian beam.  The data points were recorded for
several intensities and fixed detuning.  The solid lines represent the
(steady state) band populations in the lattice calculated \cite{PRA}
for the center-of-beam parameters using a rate equation approach.  The
experimentally measured populations and the calculated steady state
populations agree within 5\% over the full range of investigated
parameters. This is remarkable, because the calculation is based only
on the detunings and the intensities in the center of the Gaussian
beam and the comparison involves no fit parameter. Small deviations
for the experimental values towards higher ground state population are
found for high light shift parameters $U_3$. This can be attributed to
the small but finite spontaneous emission probability during the
release of the atoms. It mainly affects energetically higher lying
states and transfers additional population to the ground state.

\section{Conclusion}

\noindent 
In conclusion, we have theoretically and experimentally studied purely
optical dark optical lattices which combine low photon scattering
rates with a high population in the lowest energy bands. Long coherence
times in this bands allow for adiabatic manipulations and for the
observation of non-dissipative effects in periodic potentials
\cite{BlochOszillationen}.  The lattice is formed by coupling an
atomic ground state to two excited states. The atoms are trapped at
locations of purely circular polarization which permits an extension of
the scheme to two and three dimensions using the same field
configurations as for bright optical lattices
\cite{MultiDLattices}. The adiabatic population mapping between the
energy bands and free particle momentum intervals allows a direct
determination of the band populations. We measured the band
populations over a wide range of lattice
parameters. The quantitative agreement of the measured band
populations with the calculated results shows that adiabatic mapping
is a promising tool to study the density dependence of the band
populations in an optical lattice.

\acknowledgements 
\noindent We are grateful to Peter Marte for his stimulating and inspiring
contributions, and the Deutsche Forschungsgemeinschaft for their
support of the project.  Helmut Ritsch and Herwig Stecher wish to thank the
\"Ostereichischer Fonds zur F\"orderung der wissenschaftlichen
Forschung for their support under the grant No.~S6506/S6507.



\begin{references}
\bibitem{QuantizationAxis} Choosing the quantization
axis along the electric field vector at locations of linear
polarization simplifies the analysis. The light field then induces
$\pi$-transitions ($\Delta m=0$). At locations of purely circular
polarization the light field induces $\sigma$-transitions if the
quantization axis is chosen along the standing wave axis. These
conventions are used in Fig.~\protect\ref{Lattice}.
\end{references}
\end{document}